\documentclass{sig-alternate}
\usepackage{xcolor}
\usepackage{url}
\usepackage{verbatim} 
\usepackage{multirow}
\usepackage{algorithm}
\usepackage[noend]{algorithmic}
\usepackage{xspace}
\usepackage{graphicx}
\usepackage{epsfig}
\usepackage{amsmath}
\usepackage{amssymb}
\usepackage{times}
\usepackage{enumitem}
\usepackage[normalem]{ulem}
\usepackage{caption, subcaption}

\newcommand{\hide}[1]{}
\newcommand{\xhdr}[1]{\vspace{1.7mm}\noindent{{\bf #1.}}}

\newcommand{\eg}{\emph{e.g.}}
\newcommand{\ie}{\emph{i.e.}}

\newcommand{\nodevec}{{\textsl{node2vec}}\xspace}

\graphicspath{{./FIG/}}

\begin{document}

\title{node2vec: Scalable Feature Learning for Networks}

\numberofauthors{2}
\author{
\alignauthor
Aditya Grover\\
      \affaddr{Stanford University}\\
      \email{adityag@cs.stanford.edu}
\alignauthor Jure Leskovec\\
      \affaddr{Stanford University}\\
      \email{jure@cs.stanford.edu}
}

\CopyrightYear{2016} 
\setcopyright{acmlicensed}
\conferenceinfo{KDD '16,}{August 13 - 17, 2016, San Francisco, CA, USA}
\isbn{978-1-4503-4232-2/16/08}\acmPrice{\$15.00}
\doi{http://dx.doi.org/10.1145/2939672.2939754}

\maketitle

\begin{abstract}

Prediction tasks over nodes and edges in networks require careful effort in engineering features used by learning algorithms. Recent research in the broader field of representation learning has led to significant progress in automating prediction by learning the features themselves. However, present feature learning approaches are not expressive enough to capture the diversity of connectivity patterns observed in networks. 

Here we propose \nodevec, an algorithmic framework for learning continuous feature representations for nodes in networks. In \nodevec, we learn a mapping of nodes to a low-dimensional space of features that maximizes the likelihood of preserving network neighborhoods of nodes.
We define a flexible notion of a node's network neighborhood and design a biased random walk procedure, which efficiently explores diverse neighborhoods.
%
%
Our algorithm generalizes prior work which is based on rigid notions of network neighborhoods, and we argue that the added flexibility in exploring neighborhoods is the key to learning richer representations.

We demonstrate the efficacy of \nodevec over existing state-of-the-art techniques on multi-label classification and link prediction in several real-world networks from diverse domains.
Taken together, our work represents a new way for efficiently learning state-of-the-art task-independent representations in complex networks. 

\end{abstract}

\vspace{1mm}
\noindent {\bf Categories and Subject Descriptors:} H.2.8 {\bf [Database Management]}: Database applications---{\it Data mining}; I.2.6 {\bf [Artificial Intelligence]}: Learning

\noindent {\bf General Terms:} Algorithms; Experimentation.


\section{Introduction}
\label{sec:intro}

Many important tasks in network analysis involve predictions over nodes and edges. In a typical node classification task, we are interested in predicting the most probable labels of nodes in a network~\cite{tsoumakas2006multi}. For example, in a social network, we might be interested in predicting interests of users, or in a protein-protein interaction network we might be interested in predicting functional labels of proteins~\cite{radivojac2013large,yang-www2011}. Similarly, in link prediction, we wish to predict whether a pair of nodes in a network should have an edge connecting them~\cite{liben2007link}. Link prediction is useful in a wide variety of domains; for instance, in genomics, it helps us discover novel interactions between genes, and in social networks, it can identify real-world friends~\cite{srw,vazquez2003global}.

Any supervised machine learning algorithm requires a set of informative, discriminating, and independent features.
In prediction problems on networks this means that one has to construct a feature vector representation for the nodes and edges. A typical solution involves hand-engineering domain-specific features based on expert knowledge. Even if one discounts the tedious effort required for feature engineering, such features are usually designed for specific tasks and do not generalize across different prediction tasks.

An alternative approach is to 
 {\em learn} feature representations by solving an optimization problem~\cite{bengio-pami2013}. 
The challenge in feature learning is defining an objective function, which involves a trade-off in balancing computational efficiency and predictive accuracy. On one side of the spectrum, one could directly aim to find a feature representation that optimizes performance of a downstream prediction task. 
While this supervised procedure results in good accuracy, it comes at the cost of high training time complexity due to a blowup in the number of parameters that need to be estimated. At the other extreme, the objective function can be defined to be independent of the downstream prediction task and the representations can be learned in a purely unsupervised way. This makes the optimization computationally efficient and with a carefully designed objective, it results in task-independent features that closely match task-specific approaches in predictive accuracy~\cite{word2vec, pennington-emnlp2014}.

However, current techniques fail to satisfactorily define and optimize a reasonable objective required for scalable unsupervised feature learning in networks. Classic approaches based on linear and non-linear dimensionality reduction techniques such as Principal Component Analysis, Multi-Dimensional Scaling and their extensions~\cite{belkin-nips2001, roweis-science2000, tenenbaum-science2000, yan2007graph} optimize an objective that transforms a representative data matrix of the network such that it maximizes the variance of the data representation. Consequently, these approaches invariably involve eigendecomposition of the appropriate data matrix which is expensive for large real-world networks. Moreover, the resulting latent representations give 
poor performance on various prediction tasks over networks. 

\begin{figure}[t]
	\centering
	\includegraphics[width=0.45\textwidth]{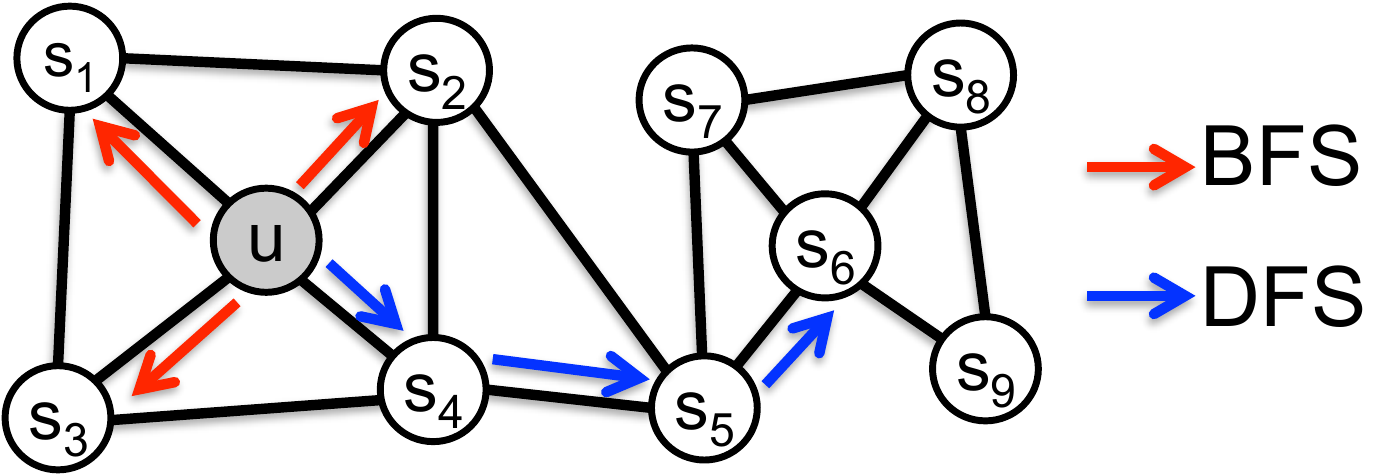}
	\caption{BFS and DFS search strategies from node $u$ ($k=3$).
	}
	\label{fig:bfsdfs}
\end{figure}

Alternatively, we can design an objective that seeks to preserve local neighborhoods of nodes. The objective can be efficiently optimized using stochastic gradient descent (SGD) akin to backpropogation on just single hidden-layer feedforward neural networks.
Recent attempts in this direction~\cite{deepwalk, line} propose 
efficient algorithms but 
rely on a rigid notion of a network neighborhood, which results in these approaches
being largely insensitive to connectivity patterns unique to networks.
Specifically, nodes in networks could be organized based on communities they belong to (\ie, \emph{homophily}); in other cases, the organization could be based on the structural roles of nodes in the network (\ie, \emph{structural equivalence})~\cite{fortunato09community,henderson-kdd2012, yang14ieee}. 
For instance, in Figure~\ref{fig:bfsdfs}, we observe nodes $u$ and $s_1$ belonging to the same tightly knit community of nodes, while the nodes $u$ and $s_6$ in the two distinct communities share the same structural role of a hub node. Real-world networks commonly exhibit a mixture of such equivalences. Thus, it is essential to allow for a flexible algorithm that can learn node representations obeying both principles: ability to learn representations that embed nodes from the same network community closely together, as well as to learn representations where nodes 
that share similar roles have similar embeddings. This would allow feature learning algorithms to generalize across a wide variety of domains and prediction tasks.

\xhdr{Present work}
We propose \nodevec, a semi-supervised algorithm for scalable feature learning in networks. We optimize
a custom graph-based objective function using SGD motivated by prior work on natural language processing 
~\cite{word2vec}.
Intuitively, our approach returns feature representations that maximize the likelihood of preserving network neighborhoods of nodes in a $d$-dimensional feature space. 
We use a 2$^{\textrm{nd}}$ order random walk approach to generate (sample) network neighborhoods for nodes.

Our key contribution is in defining a flexible notion of a node's network neighborhood. By choosing an appropriate notion of a neighborhood, \nodevec can learn representations that organize nodes based on their network roles and/or communities they belong to. We achieve this by developing a family of biased random walks, which efficiently explore diverse neighborhoods of a given node. 
The resulting algorithm is flexible, giving us control over the search space through tunable parameters, in contrast to rigid search procedures in prior work~\cite{deepwalk, line}. Consequently, our method generalizes prior work and can model the full spectrum of equivalences observed in networks. The parameters governing our search strategy have an intuitive interpretation and bias the walk towards different network exploration strategies. These parameters can also be learned directly using a tiny fraction of labeled data in a semi-supervised fashion.

We also show how feature representations of individual nodes can be extended to pairs of nodes ({\em i.e.}, edges). In order to generate feature representations of edges, we compose the learned feature representations of the individual nodes using simple binary operators. This compositionality lends \nodevec to prediction tasks involving nodes as well as edges. 

Our experiments focus on two common prediction tasks in networks: a multi-label classification task, where every node is assigned one or more class labels and a link prediction task, where we predict the existence of an edge given a pair of nodes.
We contrast the performance of \nodevec with state-of-the-art feature learning algorithms~\cite{deepwalk, line}.
We experiment with several real-world networks from diverse domains, such as social networks, information networks, as well as networks from systems biology. Experiments demonstrate that \nodevec outperforms state-of-the-art methods by up to 26.7\% on multi-label classification and up to 12.6\% on link prediction. The algorithm shows competitive performance with even 10\% labeled data and is also robust to perturbations in the form of noisy or missing edges. Computationally, the major phases of \nodevec are trivially parallelizable, and it can scale to large networks with millions of nodes in a few hours.


Overall our paper makes the following contributions:
\begin{enumerate}[noitemsep,nolistsep]
	\item We propose \nodevec,  
	an efficient scalable algorithm for feature learning in networks that efficiently optimizes a novel network-aware, neighborhood preserving objective using SGD.
	\item We show how \nodevec is in accordance with established principles in network science, providing flexibility in discovering representations conforming to different equivalences.
	\item We extend \nodevec and other feature learning methods based on neighborhood preserving objectives, from nodes to pairs of nodes for edge-based prediction tasks.
	\item We empirically evaluate \nodevec for multi-label classification and link prediction on several real-world datasets.
\end{enumerate} 

The rest of the paper is structured as follows. In Section~\ref{sec:related}, we briefly survey related work in feature learning for networks. We present the technical details for feature learning using \nodevec in Section~\ref{sec:proposed}.
In Section~\ref{sec:experiments}, we empirically evaluate \nodevec on prediction tasks over nodes and edges on various real-world networks 
and assess the parameter sensitivity, perturbation analysis, and scalability aspects of our algorithm. 
We conclude with a discussion of the \nodevec framework and highlight some promising directions for future work in Section~\ref{sec:discussion}. Datasets and a reference implementation of \nodevec are available on the project page: \url{http://snap.stanford.edu/node2vec}.

\section{Related work}
\label{sec:related}

Feature engineering has been extensively studied by the machine learning community under various headings. In networks, the conventional paradigm for generating features for nodes is based on feature extraction techniques which typically involve some seed hand-crafted features based on network properties~\cite{labelindependent,refex}. In contrast, our goal is to automate the whole process by casting feature extraction as a representation learning problem in which case we do not require any hand-engineered features. 

Unsupervised feature learning approaches typically exploit the spectral properties of various matrix representations of graphs, especially the Laplacian and the adjacency matrices. Under this linear algebra perspective, these methods can be viewed as dimensionality reduction techniques. Several linear (\eg, PCA) and non-linear (\eg, IsoMap) dimensionality reduction techniques have been proposed~\cite{belkin-nips2001, roweis-science2000, tenenbaum-science2000, yan2007graph}. These methods suffer from both computational and statistical performance drawbacks. In terms of computational efficiency, eigendecomposition of a data matrix is expensive unless the solution quality is significantly compromised with approximations, and hence, these methods are hard to scale to large networks. Secondly, these methods optimize for objectives that are not robust to the diverse patterns observed in networks (such as homophily and structural equivalence) and make assumptions about the relationship between the underlying network structure and the prediction task. For instance, spectral clustering makes a strong homophily assumption that graph cuts will be useful for classification~\cite{spectral}. Such assumptions are reasonable in many scenarios, but unsatisfactory in effectively generalizing across diverse networks.

Recent advancements in representational learning for natural language processing opened new ways for feature learning of discrete objects such as words.
In particular, the Skip-gram model~\cite{word2vec} aims to learn continuous feature representations for words by optimizing a neighborhood preserving likelihood objective. 
The algorithm proceeds as follows: It scans over the words of a document, and for every word it aims to embed it such that the word's features can predict nearby words (\ie, words inside some context window). The word feature representations are learned by optmizing the likelihood objective using SGD with negative sampling~\cite{word2vec2}. 
%
The Skip-gram objective is based on the distributional hypothesis which states that words in similar contexts tend to have similar meanings~\cite{harris-1954}. That is, similar words tend to appear in similar word neighborhoods.

Inspired by the Skip-gram model, recent research established an analogy for networks by representing a network as a ``document''~\cite{deepwalk, line}.
The same way as a document is an ordered sequence of words, one could sample sequences of nodes from the underlying network and turn a network into a ordered sequence of nodes. However, there are many possible sampling strategies for nodes, resulting in different learned feature representations. In fact, as we shall show, there is no clear winning sampling strategy that works across all networks and all prediction tasks. This is a major shortcoming of prior work which fail to offer any flexibility in sampling of nodes from a network~\cite{deepwalk, line}. Our algorithm \nodevec overcomes this limitation by designing a flexible objective that is not tied to a particular sampling strategy and provides parameters to tune the explored search space (see Section~\ref{sec:proposed}).



Finally, for both node and edge based prediction tasks, there is a body of recent work for supervised feature learning based on existing and novel graph-specific deep network architectures~\cite{li-icdm2014, li-icdm2014-2, li2015gated, tian-aaai2014, zhai-sdm2015}. These architectures directly minimize the loss function for a downstream prediction task using several layers of non-linear transformations which results in high accuracy, but at the cost of scalability due to high training time requirements.


\section{Feature learning framework}
\label{sec:proposed}

We formulate feature learning in networks as a maximum likelihood optimization problem. Let $G = (V, E)$ be a given network. Our analysis is general and applies to any (un)directed, (un)weighted network. 
Let $f: V \rightarrow \mathbb{R}^d$ be the mapping function from nodes to feature representaions we aim to learn for a downstream prediction task. Here $d$ is a parameter specifying the number of dimensions of our feature representation. 
Equivalently, $f$ is a matrix of size $|V| \times d$ parameters. 
For every source node $u \in V$, we define $N_S(u) \subset V$ as a {\em network neighborhood} of node $u$ generated through a neighborhood sampling strategy $S$. 

We proceed by extending the Skip-gram architecture to networks \cite{word2vec,deepwalk}. 
We seek to optimize the following objective function, which maximizes the log-probability of observing a network neighborhood $N_S(u)$ for a node $u$ conditioned on its feature representation, given by $f$:
\vspace{-0.05in}
\begin{equation}\label{eq:rawObjFn}
\begin{aligned}
& \underset{f}{\text{max}}
& &  \sum_{u \in V}\log Pr(N_S(u)| f(u)).
\end{aligned}
\end{equation}

In order to make the optimization problem tractable, we make two standard assumptions:
\begin{itemize}[noitemsep,nolistsep]
	\item Conditional independence. We factorize the likelihood by assuming that the likelihood of observing a neighborhood node is independent of observing any other neighborhood node given the feature representation of the source:
	\begin{align}
		Pr(N_S(u)| f(u)) = \prod_{n_i \in N_S(u)} Pr(n_i | f(u)). \nonumber
	\end{align}
	\item Symmetry in feature space. A source node and neighborhood node have a symmetric effect over each other in feature space. Accordingly, we model the conditional likelihood of every source-neighborhood node pair as a softmax unit parametrized by a dot product of their features:

	\begin{align}
	Pr(n_i | f(u)) = \frac{\exp(f(n_i)\cdot f(u))}{\sum_{v \in V} \exp(f(v)\cdot f(u))}. \nonumber
	\end{align}
\end{itemize}
With the above assumptions, the objective in Eq.~\ref{eq:rawObjFn} simplifies to:
\vspace{-0.05in}
\begin{equation}\label{eq:finalObjFn}
	\begin{aligned}
	& \underset{f}{\text{max}}
	& &    \sum_{u \in V} \bigg[- \log Z_u + \sum_{n_i \in N_S(u)} f(n_i)\cdot f(u)\bigg].
	\end{aligned}
\end{equation}
The per-node partition function, $Z_u=\sum_{v \in V} \exp(f(u)\cdot f(v))$, is expensive to compute for large networks and we approximate it using negative sampling~\cite{word2vec2}. We optimize Eq.~\ref{eq:finalObjFn} using stochastic gradient ascent over the model parameters defining the features $f$.

Feature learning methods based on the Skip-gram architecture have been originally developed in the context of natural language~\cite{word2vec}. Given the linear nature of text, the notion of a neighborhood can be naturally defined using a sliding window over consecutive words. Networks, however, are not linear, and thus a richer notion of a neighborhood is needed.
To resolve this issue, we propose a randomized procedure that samples many different neighborhoods of a given source node $u$. The neighborhoods $N_S(u)$ are not restricted to just immediate neighbors but can have vastly different structures depending on the sampling strategy $S$. 



\subsection{Classic search strategies}
\label{sec:bfsdfs}



We view the problem of sampling neighborhoods of a source node as a form of local search. Figure \ref{fig:bfsdfs} shows a graph, where given a source node $u$ we aim to generate (sample) its neighborhood $N_S(u)$. Importantly, to be able to fairly compare different sampling strategies $S$, we shall constrain the size of the neighborhood set $N_S$ to $k$ nodes and then sample multiple sets for a single node $u$. Generally, there are two extreme sampling strategies for generating neighborhood set(s) $N_S$ of $k$ nodes:
\begin{itemize}[noitemsep,nolistsep]
	\item \textbf {Breadth-first Sampling (BFS)} The neighborhood $N_S$ is restricted to nodes which are immediate neighbors of the source. For example, in Figure~\ref{fig:bfsdfs} for a neighborhood of size $k=3$, BFS samples nodes $s_1$, $s_2$, $s_{3}$.
	\item \textbf {Depth-first Sampling (DFS)} The neighborhood consists of nodes sequentially sampled at increasing distances from the source node. In Figure~\ref{fig:bfsdfs}, DFS samples $s_{4}$, $s_{5}$, $s_{6}$. 
\end{itemize}



The breadth-first and depth-first sampling represent extreme scenarios in terms of the search space they explore leading to interesting implications on the learned representations. 

In particular, prediction tasks on nodes in networks often shuttle between two kinds of similarities: homophily and structural equivalence~\cite{latentfactor}. Under the homophily hypothesis~\cite{fortunato09community, yang14ieee} nodes that are highly interconnected and belong to similar network clusters or communities should be embedded closely together (\eg, nodes $s_1$ and $u$ in Figure~\ref{fig:bfsdfs} belong to the same network community). In contrast, under the structural equivalence hypothesis~\cite{henderson-kdd2012} nodes that have similar structural roles in networks should be embedded closely together (\eg, nodes $u$ and $s_6$ in Figure~\ref{fig:bfsdfs} act as hubs of their corresponding communities). Importantly, unlike homophily, structural equivalence does not emphasize connectivity; nodes could be far apart in the network and still have the same structural role.
In real-world, these equivalence notions are not exclusive; networks commonly exhibit both behaviors where some nodes exhibit homophily while others reflect structural equivalence.

We observe that BFS and DFS strategies play a key role in producing representations that reflect either of the above equivalences. In particular, the neighborhoods sampled by BFS lead to embeddings that correspond closely to structural equivalence. Intuitively, we note that in order to ascertain structural equivalence, it is often sufficient to characterize the local neighborhoods accurately. For example, structural equivalence based on network roles such as bridges and hubs can be inferred just by observing the immediate neighborhoods of each node. By restricting search to nearby nodes, BFS achieves this characterization and obtains a microscopic view of the neighborhood of every node. Additionally, in BFS, nodes in the sampled neighborhoods tend to repeat many times. This is also important as it reduces the variance in characterizing the distribution of 1-hop nodes with respect the source node. However, a very small portion of the graph is explored for any given $k$.

The opposite is true for DFS which can explore larger parts of the network as it can move further away from the source node $u$ (with sample size $k$ being fixed). In DFS, the sampled nodes more accurately reflect a macro-view of the neighborhood which is essential in inferring communities based on homophily. However, the issue with DFS is that it is important to not only infer which node-to-node dependencies exist in a network, but also to characterize the exact nature of these dependencies. This is hard given we have a constrain on the sample size and a large neighborhood to explore, resulting in high variance. Secondly, moving to much greater depths leads to complex dependencies since a sampled node may be far from the source and potentially less representative. 

\subsection{node2vec}

Building on the above observations, we design a flexible neighborhood sampling strategy which allows us to smoothly interpolate between BFS and DFS. We achieve this by developing a flexible biased random walk procedure that can explore neighborhoods in a BFS as well as DFS fashion.

\subsubsection{Random Walks}
Formally, given a source node $u$, we simulate a random walk of fixed length $l$. Let $c_i$ denote the $i${th} node in the walk, starting with $c_0 = u$. Nodes $c_i$ are generated by the following distribution:
\[
	P(c_i = x \mid c_{i-1} = v) =
	\begin{cases}
	\frac{\pi_{vx}}{Z} & \text{if } (v,x) \in E \\
	0 & \text{otherwise}
	\end{cases}
\]
where $\pi_{vx}$ is the unnormalized transition probability between nodes $v$ and $x$, and $Z$ is the normalizing constant. 

\subsubsection{Search bias \large{$\alpha$}}
The simplest way to bias our random walks would be to sample the next node based on the static edge weights $w_{vx}$ \ie, $\pi_{vx} = w_{vx}$. (In case of unweighted graphs $w_{vx}=1$.) However, this does not allow us to account for the network structure and guide our search procedure to explore different types of network neighborhoods. 
Additionally, unlike BFS and DFS which are extreme sampling paradigms suited for structural equivalence and homophily respectively, our random walks should accommodate for the fact that these notions of equivalence are not competing or exclusive, and real-world networks commonly exhibit a mixture of both. 

\begin{figure}[t]
	\centering
	\includegraphics[width=0.3\textwidth]{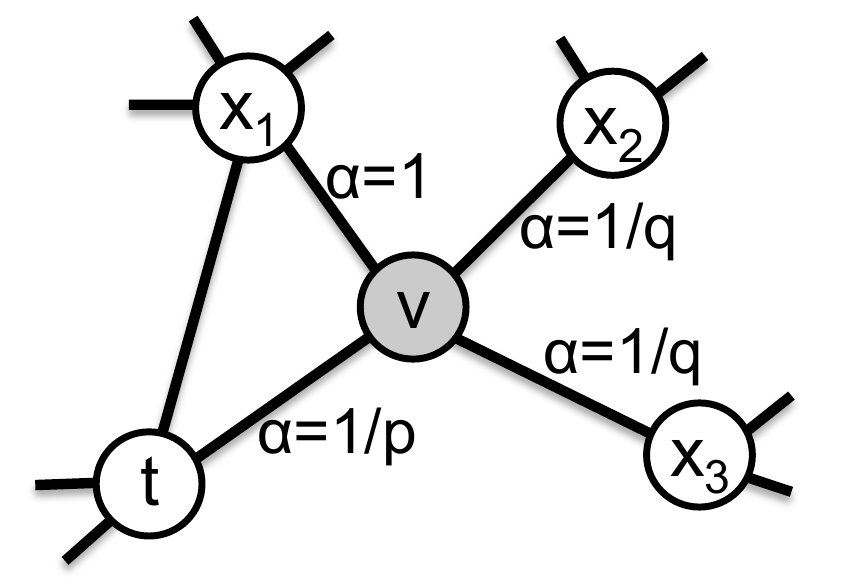}
	\vspace{-0.2cm}
	\caption{Illustration of the random walk procedure in \nodevec. The walk just transitioned from $t$ to $v$ and
	is now evaluating its next step out of node $v$. Edge labels indicate search biases $\alpha$.}
	\label{fig:walk}
	\vspace{-0.4cm}
\end{figure}

We define a 2$^{\textrm{nd}}$ order random walk with two parameters $p$ and $q$ which guide the walk: Consider a random walk that just traversed edge $(t,v)$ and now resides at node $v$ (Figure~\ref{fig:walk}). The walk now needs to decide on the next step so it evaluates the transition probabilities $\pi_{vx}$ on edges $(v,x)$ leading from $v$. We set the unnormalized transition probability to $\pi_{vx} = \alpha_{pq}(t,x)\cdot w_{vx}$, where 
\[
	\alpha_{pq}(t,x) = 
	\begin{cases}
	\frac{1}{p}  & \text{if } d_{tx} = 0\\
	1 & \text{if } d_{tx} = 1\\
	\frac{1}{q} & \text{if } d_{tx} = 2
	\end{cases}
\]
and $d_{tx}$ denotes the shortest path distance between nodes $t$ and $x$. Note that $d_{tx}$ must be one of $\{0,1,2\}$, and hence, the two parameters are necessary and sufficient to guide the walk.

Intuitively, parameters $p$ and $q$ control how fast the walk explores and leaves the neighborhood of starting node $u$. In particular, the parameters allow our search procedure to (approximately) interpolate between BFS and DFS and thereby reflect an affinity for different notions of node equivalences.


\xhdr{Return parameter, $p$} Parameter $p$ controls the likelihood of immediately revisiting a node in the walk. Setting it to a high value ($>\max(q, 1)$) ensures that we are less likely to sample an already-visited node in the following two steps (unless the next node in the walk had no other neighbor). This strategy encourages moderate exploration and avoids $2$-hop redundancy in sampling. On the other hand, if $p$ is low ($<\min(q, 1)$), it would lead the walk to backtrack a step (Figure~\ref{fig:walk}) and this would keep the walk ``local'' close to the starting node $u$.


\xhdr{In-out parameter, $q$} Parameter $q$ allows the search to differentiate between ``inward'' and ``outward'' nodes.  Going back to Figure~\ref{fig:walk}, if $q>1$, the random walk is biased towards nodes close to node $t$. Such walks obtain a local view of the underlying graph with respect to the start node in the walk and approximate BFS behavior in the sense that our samples comprise of nodes within a small locality.  

In contrast, if $q<1$, the walk is more inclined to visit nodes which are further away from the node $t$. Such behavior is reflective of DFS which encourages outward exploration. However, an essential difference here is that we achieve DFS-like exploration within the random walk framework. Hence, the sampled nodes are not at strictly increasing distances from a given source node $u$, but in turn, we benefit from tractable preprocessing and superior sampling efficiency of random walks. Note that by setting $\pi_{v,x}$ to be a function of the preceeding node in the walk $t$, the random walks are 2$^{\textrm{nd}}$ order Markovian.

\xhdr{Benefits of random walks} There are several benefits of random walks over pure BFS/DFS approaches. Random walks are computationally efficient in terms of both space and time requirements. The space complexity to store the immediate neighbors of every node in the graph is $O(|E|)$. 
For 2$^{\textrm{nd}}$ order random walks, it is helpful to store the interconnections between the neighbors of every node, which incurs a space complexity of $O(a^2|V|)$ where $a$ is the average degree of the graph and is usually small for real-world networks. 
The other key advantage of random walks over classic search-based sampling strategies is its time complexity. 
In particular, by imposing graph connectivity in the sample generation process, random walks provide a convenient mechanism to increase the effective sampling rate by reusing samples across different source nodes. By simulating a random walk of length $l>k$ we can generate $k$ samples for $l-k$ nodes at once due to the Markovian nature of the random walk. Hence, our effective complexity is $O\big(\frac{l}{k(l-k)}\big)$ per sample. For example, in Figure \ref{fig:bfsdfs} we sample a random walk $\{u, s_4, s_5, s_6, s_8, s_9\}$ of length $l=6$, which results in $N_S(u) =\{s_4, s_5, s_6\}$, $N_S(s_4)=\{s_5, s_6, s_8\}$ and $N_S(s_5)=\{s_6, s_8, s_9\}$.
Note that sample reuse can introduce some bias in the overall procedure. However, we observe that it greatly improves the efficiency.


\subsubsection{The \nodevec algorithm}
 
\begin{algorithm}[h]
\begin{algorithmic}
\STATE \hspace{-0.15in}{\bf LearnFeatures}{ (Graph $G=(V, E, W)$, Dimensions $d$, Walks per node $r$, Walk length $l$, Context size $k$, Return $p$, In-out $q$)}
\STATE $\pi =$ PreprocessModifiedWeights($G, p, q$)
\STATE $G' = (V, E, \pi)$
\STATE Initialize $walks$ to Empty
\FOR {$iter = 1$ {\bfseries to} $r$} 
\FORALL {nodes $u \in V$}
\STATE $walk =$ node2vecWalk($G', u, l$)
\STATE Append $walk$ to $walks$
\ENDFOR
\ENDFOR
\STATE $f =$ StochasticGradientDescent($k$, $d$, $walks$)
\STATE {\bf return} $f$
\STATE \hspace{-0.15in} \hrulefill
\STATE \hspace{-0.15in}{\bf node2vecWalk}{ (Graph $G'= (V, E, \pi)$, Start node $u$, Length $l$)}
\STATE Inititalize $walk$ to $[u]$
\FOR {$walk\_iter = 1$ {\bfseries to} $l$}
\STATE $curr = walk[-1]$
\STATE $V_{curr} =$ GetNeighbors($curr$, $G'$)
\STATE $s =$ AliasSample($V_{curr}, \pi$)
\STATE Append $s$ to $walk$
\ENDFOR
\STATE {\bf return} $walk$
\end{algorithmic}
\caption{The \nodevec algorithm.}
\label{alg:node2vec}
\end{algorithm}

The pseudocode for \nodevec, is given in Algorithm~\ref{alg:node2vec}. In any random walk, there is an implicit bias due to the choice of the start node $u$. Since we learn representations for all nodes, we offset this bias by simulating $r$ random walks of fixed length $l$ starting from {\em every} node. At every step of the walk, sampling is done based on the transition probabilities $\pi_{vx}$.
The transition probabilities $\pi_{vx}$ for the 2$^{\textrm{nd}}$ order Markov chain can be precomputed and hence, sampling of nodes while simulating the random walk can be done efficiently in $O(1)$ time using alias sampling. The three phases of \nodevec, \ie, preprocessing to compute transition probabilities, random walk simulations and optimization using SGD, are executed sequentially. Each phase is parallelizable and executed asynchronously, contributing to the overall scalability of \nodevec.

\nodevec is available at: \url{http://snap.stanford.edu/node2vec}. 

\begin{table}
\centering
\begin{tabular}{l|c|c}
\textbf{Operator} & \textbf{Symbol} & \textbf{Definition} \\ \hline
Average & $\boxplus$ & $[f(u) \boxplus f(v)]_i = \frac{f_i(u)+f_i(v)}{2}$ \\
Hadamard & $\boxdot$ & $[f(u) \boxdot f(v)]_i = f_i(u)*f_i(v)$ \\
Weighted-L1 & $\| \cdot \|_{\bar{1}}$ & $\|f(u) \cdot f(v)\|_{\bar{1}i} = |f_i(u)-f_i(v)|$ \\
Weighted-L2 & $\| \cdot \|_{\bar{2}}$ & $\|f(u) \cdot f(v)\|_{\bar{2}i} = |f_i(u)-f_i(v)|^2$
\end{tabular}
\vspace{-0.2cm}
\caption{Choice of binary operators $\circ$ for learning edge features. The definitions correspond to the $i${th} component of $g(u,v)$.
}
\vspace{-0.4cm}
\label{tab:edgeop}
\end{table} 

\subsection{Learning edge features}
The \nodevec algorithm provides a semi-supervised method to learn rich feature representations for nodes in a network. However, we are often interested in prediction tasks involving pairs of nodes instead of individual nodes. For instance, in link prediction, we predict whether a link exists between two nodes in a network. Since our random walks are naturally based on the connectivity structure between nodes in the underlying network, we extend them to pairs of nodes using a bootstrapping approach over the feature representations of the individual nodes.

Given two nodes $u$ and $v$, we define a binary operator $\circ$ over the corresponding feature vectors $f(u)$ and $f(v)$ in order to generate a representation $g(u,v)$ such that $g: V \times V \rightarrow \mathbb{R}^{d'}$ where $d'$ is the representation size for the pair $(u,v)$. We want our operators to be generally defined for any pair of nodes, even if an edge does not exist between the pair since doing so makes the representations useful for link prediction where our test set contains both true and false edges (\ie, do not exist). We consider several choices for the operator $\circ$ such that $d'= d$ which are summarized in Table \ref{tab:edgeop}.

\section{Experiments}
\label{sec:experiments}

The objective in Eq.~\ref{eq:finalObjFn} is independent of any downstream task and the flexibility in exploration offered by \nodevec lends the learned feature representations to a wide variety of network analysis settings discussed below.

\subsection{Case Study: Les Mis\'{e}rables network}

In Section~\ref{sec:bfsdfs} we observed that BFS and DFS strategies represent extreme ends on the spectrum of embedding nodes based on the principles of homophily (\ie, network communities) and structural equivalence (\ie, structural roles of nodes). We now aim to empirically demonstrate this fact and show that \nodevec in fact can discover embeddings that obey both principles.

We use a network where nodes correspond to characters in the novel Les Mis\'{e}rables~\cite{knuth-1993} and edges connect coappearing characters. The network has 77 nodes and 254 edges. We set $d=16$ and run \nodevec to learn feature representation for every node in the network. The feature representations are clustered using $k$-means. We then visualize the original network in two dimensions with nodes now assigned colors based on their clusters.

\begin{figure}[t]
	\centering
	\includegraphics[width=0.4\textwidth]{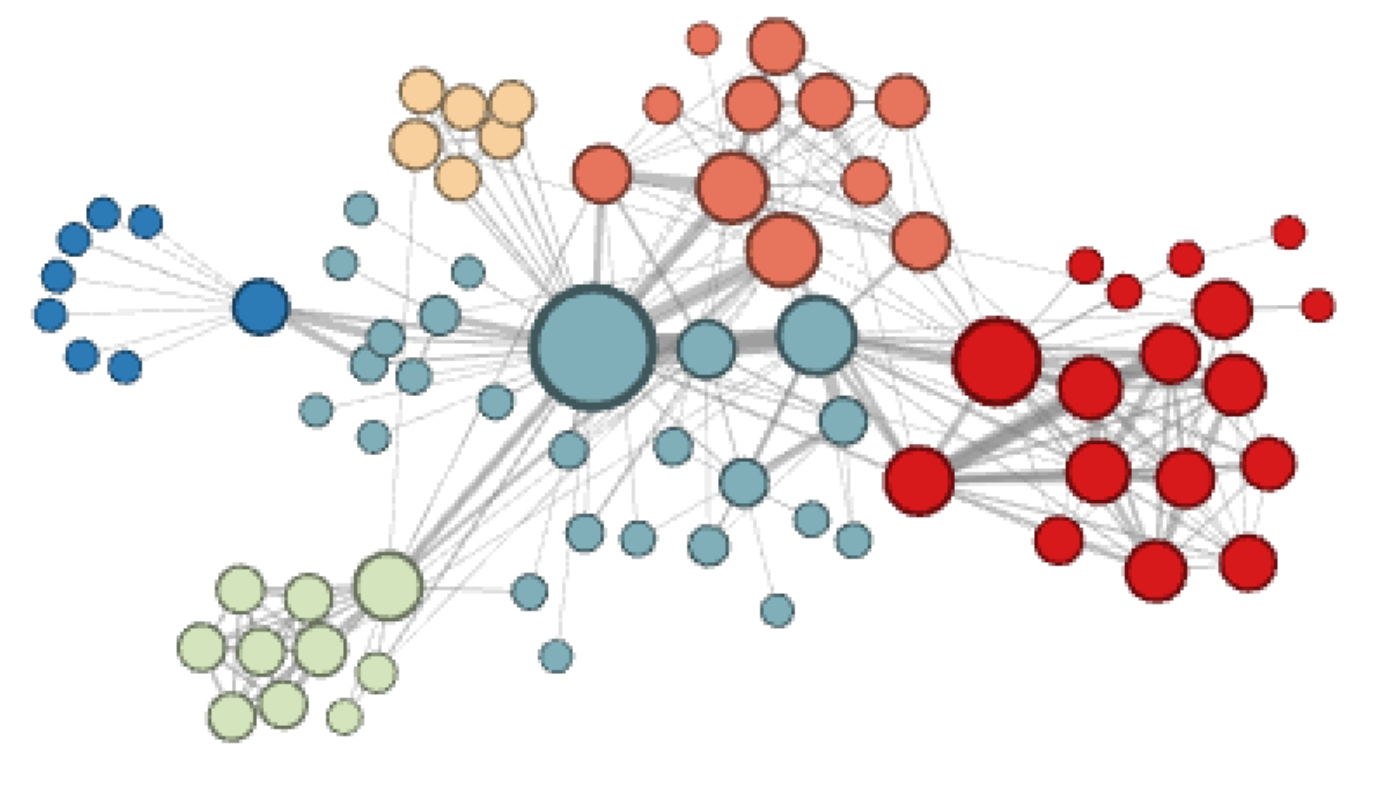}
	\vspace{-0.2cm}
	\includegraphics[width=0.4\textwidth]{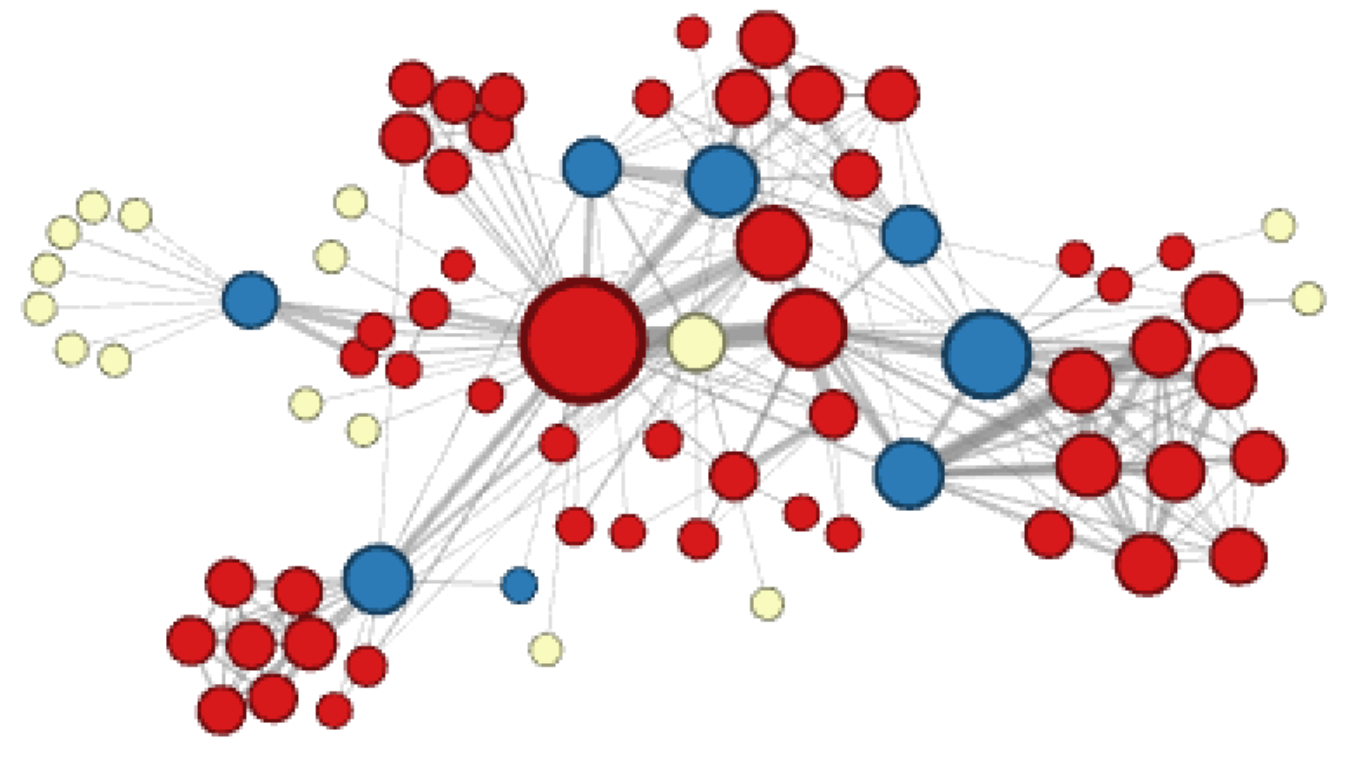}
	\caption{Complementary visualizations of Les Mis\'{e}rables coappearance network generated by \nodevec with label colors reflecting homophily (top) and structural equivalence (bottom).}\label{fig:equi}
	\vspace{-0.4cm}
\end{figure}

Figure~\ref{fig:equi}(top) shows the example when we set $p=1, q=0.5$. Notice how regions of the network (\ie, network communities) are colored using the same color. In this setting \nodevec discovers clusters/communities of characters that frequently interact with each other in the major sub-plots of the novel. Since the edges between characters are based on coappearances, we can conclude this characterization closely relates with homophily. 

In order to discover which nodes have the same structural roles we use the same network but set $p=1, q=2$, use \nodevec to get node features and then cluster the nodes based on the obtained features. Here \nodevec obtains a complementary assignment of node to clusters such that the colors correspond to structural equivalence as illustrated in Figure~\ref{fig:equi}(bottom). For instance, \nodevec embeds blue-colored nodes close together. These nodes represent characters that act as bridges between different sub-plots of the novel. Similarly, the yellow nodes mostly represent characters that are at the periphery and have limited interactions. 
One could assign alternate semantic interpretations to these clusters of nodes, but the key takeaway is that \nodevec is not tied to a particular notion of equivalence. As we show through our experiments, these equivalence notions are commonly exhibited in most real-world networks and have a significant impact on the performance of the learned representations for prediction tasks.

\subsection{Experimental setup}

Our experiments evaluate the feature representations obtained through \nodevec on standard supervised learning tasks: multi-label classification for nodes and link prediction for edges. For both tasks, we evaluate the performance of \nodevec against the following feature learning algorithms:

\begin{itemize}[noitemsep,nolistsep]
	\item Spectral clustering~\cite{spectral}: This is a matrix factorization approach in which we take the top $d$ eigenvectors of the normalized Laplacian matrix of graph $G$ as the feature vector representations for nodes.
	\item DeepWalk 
	~\cite{deepwalk}: This approach learns $d$-dimensional feature representations by simulating uniform random walks. The sampling strategy in DeepWalk can be seen as a special case of \nodevec with $p=1$ and $q=1$.
	\item LINE 
	~\cite{line}: This approach learns $d$-dimensional feature representations in two separate phases. In the first phase, it learns $d/2$ dimensions by BFS-style simulations over immediate neighbors of nodes. In the second phase, it learns the next $d/2$ dimensions by sampling nodes strictly at a 2-hop distance from the source nodes. 
\end{itemize}

We exclude other matrix factorization approaches which have already been shown to be inferior 
to DeepWalk~\cite{deepwalk}. We also exclude a recent approach, GraRep~\cite{grarep}, that generalizes LINE to incorporate information from network neighborhoods beyond 2-hops, but is unable to efficiently scale to large networks. 

In contrast to the setup used in prior work for evaluating sampling-based feature learning algorithms, we generate an equal number of samples for each method and then evaluate the quality of the obtained features on the prediction task. In doing so, we discount for performance gain observed purely because of the implementation language (C/C++/Python) since it is secondary to the algorithm.
Thus, in the sampling phase, the parameters for DeepWalk, LINE and \nodevec are set such that they generate equal number of samples at runtime. As an example, if $\mathcal{K}$ is the overall sampling budget, then the \nodevec parameters satisfy $\mathcal{K}=r\cdot l \cdot |V|$. In the optimization phase, all these benchmarks optimize using SGD with two key differences that we correct for. First, DeepWalk uses hierarchical sampling to approximate the softmax probabilities with an objective similar to the one use by \nodevec. However, hierarchical softmax is inefficient when compared with negative sampling~\cite{word2vec2}. Hence, keeping everything else the same, we switch to negative sampling in DeepWalk which is also the de facto approximation in \nodevec and LINE. Second, both \nodevec and DeepWalk have a parameter for the number of context neighborhood nodes to optimize for and the greater the number, the more rounds of optimization are required. This parameter is set to unity for LINE, but since LINE completes a single epoch quicker than other approaches, we let it run for $k$ epochs.

The parameter settings used for \nodevec are in line with typical values used for DeepWalk and LINE. Specifically, we set $d=128$, $r=10$, $l=80$, $k=10$, and the optimization is run for a single epoch. We repeat our experiments for $10$ random seed initializations, and our results are statistically significant with a p-value of less than 0.01.
The best in-out and return hyperparameters were learned using 10-fold cross-validation on 10\% labeled data with a grid search over $p, q \in \{0.25, 0.50, 1, 2, 4 \}$. 

\subsection{Multi-label classification}

\begin{table}[t]
	\centering
	
	\begin{tabular}{l|c|c|c}
	\multicolumn{1}{c}{\textbf{Algorithm}} & \multicolumn{3}{c}{\textbf{Dataset}}      \\ 
										   & BlogCatalog     & PPI             & Wikipedia   \\ \hline
	Spectral Clustering					   & 0.0405			 & 0.0681	 	   & 0.0395		\\
	DeepWalk                               & 0.2110		     & 0.1768          & 0.1274 \\
	LINE                                   & 0.0784          & 0.1447          & 0.1164    \\
	\nodevec 			                   & \textbf{0.2581} & \textbf{0.1791} & \textbf{0.1552}     \\ \hline
	\nodevec settings (p,q)		  		   & 0.25, 0.25  	 & 4, 1	       		& 4, 0.5		\\
	{\bf Gain of \nodevec [\%]}         	& {\bf 22.3}	 & {\bf 1.3}       & {\bf 21.8 }
	\end{tabular}
	\vspace{-0.2cm}
	\caption{Macro-F$_1$ scores for multilabel classification on BlogCatalog, PPI (Homo sapiens) and Wikipedia word cooccurrence networks with 50\% of the nodes labeled for
	training.}
\label{tab:mlc_macro}
\vspace{-0.4cm}
\end{table}

\begin{figure*}[t]
\centering
	\includegraphics[width=\textwidth]{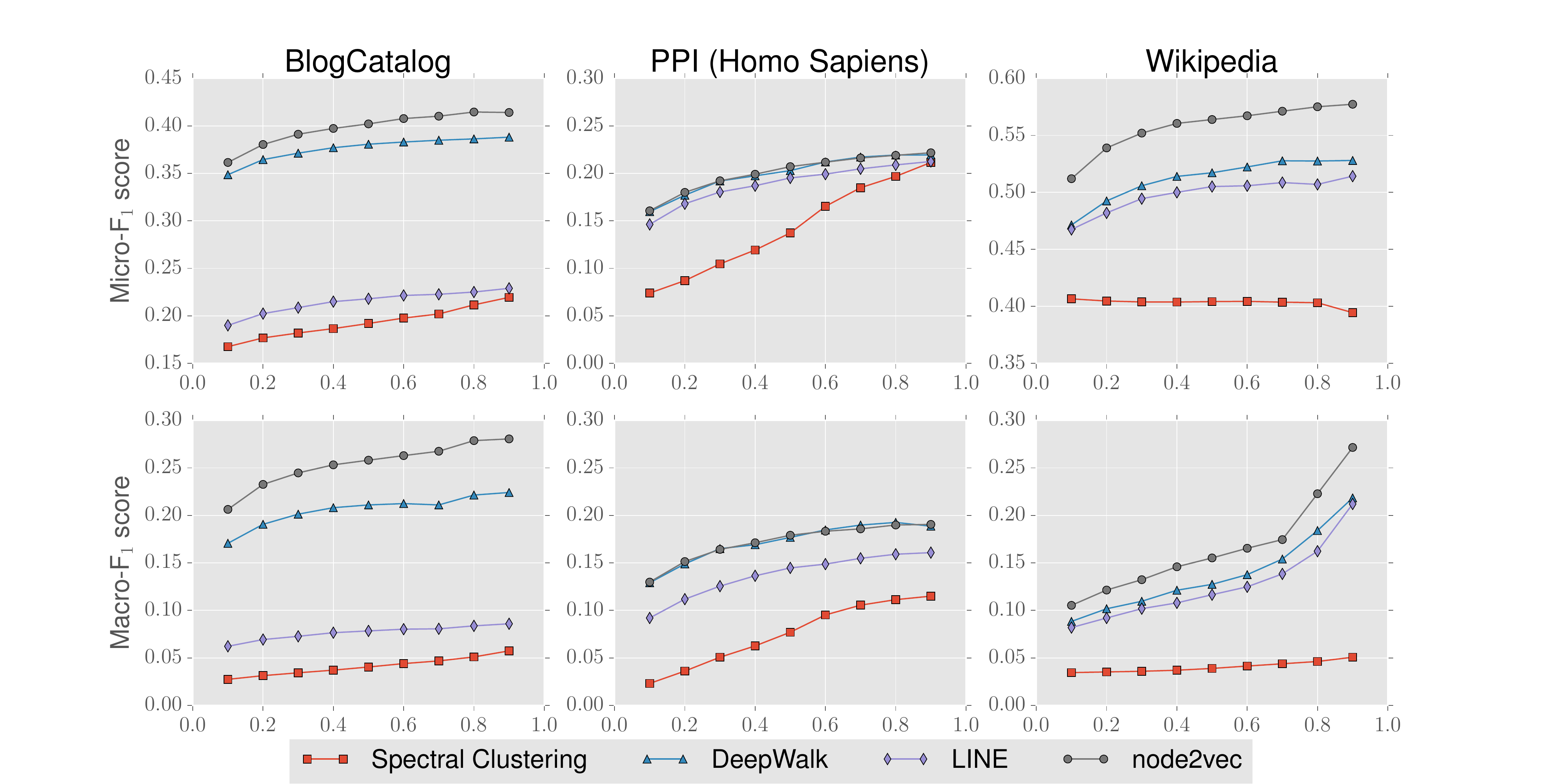}
	\vspace{-0.3cm}
	\caption{Performance evaluation of different benchmarks on varying the amount of labeled data used for training. The $x$ axis denotes the fraction of labeled data, whereas the $y$ axis in the top and bottom rows denote the Micro-F$_1$ and Macro-F$_1$ scores respectively. DeepWalk and \nodevec give comparable performance on PPI. In all other networks, across all fractions of labeled data \nodevec performs best.}\label{fig:mlc}
\vspace{-0.4cm}
\end{figure*}

In the multi-label classification setting, every node is assigned one or more labels from a finite set $\mathcal{L}$. During the training phase, we observe a certain fraction of nodes and all their labels. The task is to predict the labels for the remaining nodes. This is a challenging task especially if $\mathcal{L}$ is large. 
We utilize the following datasets:
\begin{itemize}[noitemsep,nolistsep]
	\item BlogCatalog~\cite{asu}: This is a network of social relationships of the bloggers listed on the BlogCatalog website. The labels represent blogger interests inferred through the meta-data provided by the bloggers. The network has 10,312 nodes, 333,983 edges, and 39 different labels.

	\item Protein-Protein Interactions (PPI)~\cite{biogrid}: We use a subgraph of the PPI network for Homo Sapiens. The subgraph corresponds to the graph induced by nodes for which we could obtain labels from the hallmark gene sets~\cite{msigdb} and represent biological states. 
	The network has 3,890 nodes, 76,584 edges, and 50 different labels.

	\item Wikipedia ~\cite{wiki-pos}: This is a cooccurrence network of words appearing in the first million bytes of the Wikipedia dump. The labels represent the Part-of-Speech (POS) tags inferred using the Stanford POS-Tagger~\cite{postagger}. The network has 4,777 nodes, 184,812 edges, and 40 different labels.
\end{itemize}

All these networks exhibit a fair mix of homophilic and structural equivalences. For example, we expect the social network of bloggers to exhibit strong homophily-based relationships; however, there might also be some ``familiar strangers'', \ie, bloggers that do not interact but share interests and hence are structurally equivalent nodes.
The biological states of proteins in a protein-protein interaction network also exhibit both types of equivalences. For example, they exhibit structural equivalence when proteins perform functions complementary to those of neighboring proteins, and at other times, they organize based on homophily in assisting neighboring proteins in performing similar functions. The word cooccurence network is fairly dense, since edges exist between words cooccuring in a 2-length window in the Wikipedia corpus. Hence, words having the same POS tags are not hard to find, lending a high degree of homophily. At the same time, we expect some structural equivalence in the POS tags due to syntactic grammar patterns such as nouns following determiners, punctuations succeeding nouns etc.

\xhdr{Experimental results} 
The node feature representations are input to a one-vs-rest logistic regression classifier with L2 regularization. The train and test data is split equally over 10 random instances. We use the Macro-F$_1$ scores for comparing performance in Table~\ref{tab:mlc_macro} and the relative performance gain is over the closest benchmark. The trends are similar for Micro-F$_1$ and accuracy and are not shown. 

From the results, it is evident we can see how the added flexibility in exploring neighborhoods allows \nodevec to outperform the other benchmark algorithms. In BlogCatalog, we can discover the right mix of homophily and structural equivalence by setting parameters $p$ and $q$ to low values, giving us 22.3\% gain over DeepWalk and 229.2\% gain over LINE in Macro-F$_1$ scores. LINE showed worse performance than expected, which can be explained by its inability to reuse samples, a feat that can be easily done using the random walk methods. Even in our other two networks, where we have a mix of equivalences present, the semi-supervised nature of \nodevec can help us infer the appropriate degree of exploration necessary for feature learning. In the case of PPI network, the best exploration strategy ($p=4$, $q=1$) turns out to be virtually indistinguishable from DeepWalk's uniform ($p=1$, $q=1$) exploration giving us only a slight edge over DeepWalk by avoiding redudancy in already visited nodes through a high $p$ value, but a convincing 23.8\% gain over LINE in Macro-F$_1$ scores. However, in general, the uniform random walks can be much worse than the exploration strategy learned by \nodevec. As we can see in the Wikipedia word cooccurrence network, uniform walks cannot guide the search procedure towards the best samples and hence, we achieve a gain of 21.8\% over DeepWalk and 33.2\% over LINE.

For a more fine-grained analysis, we also compare performance while varying the train-test split from 10\% to 90\%, while learning parameters $p$ and $q$ on 10\% of the data as before. For brevity, we summarize the results for the Micro-F$_1$ and Macro-F$_1$ scores graphically in Figure~\ref{fig:mlc}. Here we make similar observations. All methods significantly outperform Spectral clustering, DeepWalk outperforms LINE, \nodevec consistently outperforms LINE and achieves large improvement over DeepWalk across domains. For example, we achieve the biggest improvement over DeepWalk of 26.7\% on BlogCatalog at 70\% labeled data. In the worst case, the search phase has little bearing on learned representations in which case \nodevec is equivalent to DeepWalk. Similarly, the improvements are even more striking when compared to LINE, where in addition to drastic gain (over 200\%) on BlogCatalog, we observe high magnitude improvements upto 41.1\% on other datasets such as PPI while training on just 10\% labeled data.


\begin{figure*}[t]
\centering
\begin{subfigure}[b]{0.70\textwidth}
\includegraphics[width=\textwidth]{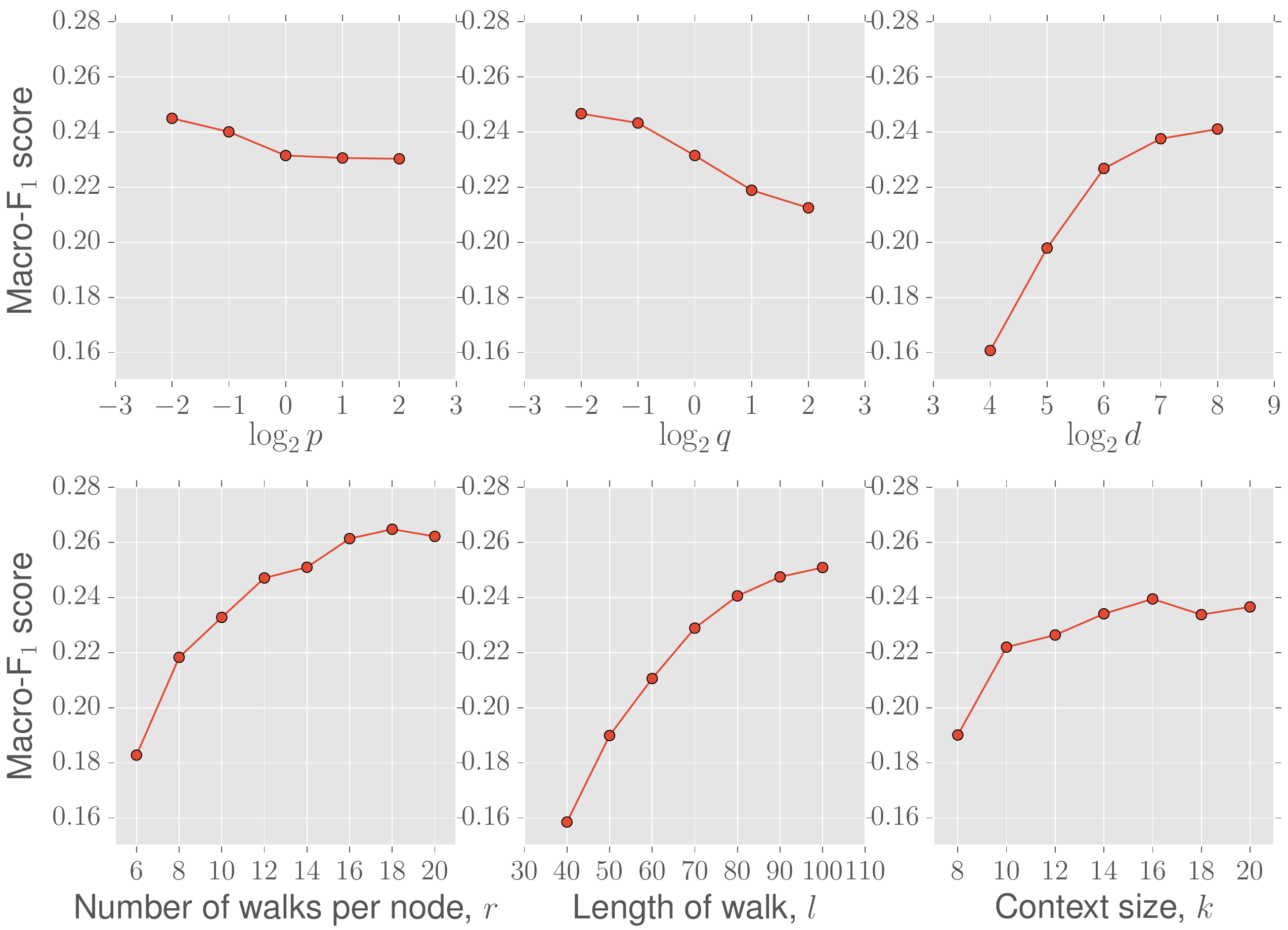}
\caption{}\label{fig:mlc_params}
\end{subfigure} 
~
\begin{subfigure}[b]{0.28\textwidth}
\includegraphics[width=\textwidth]{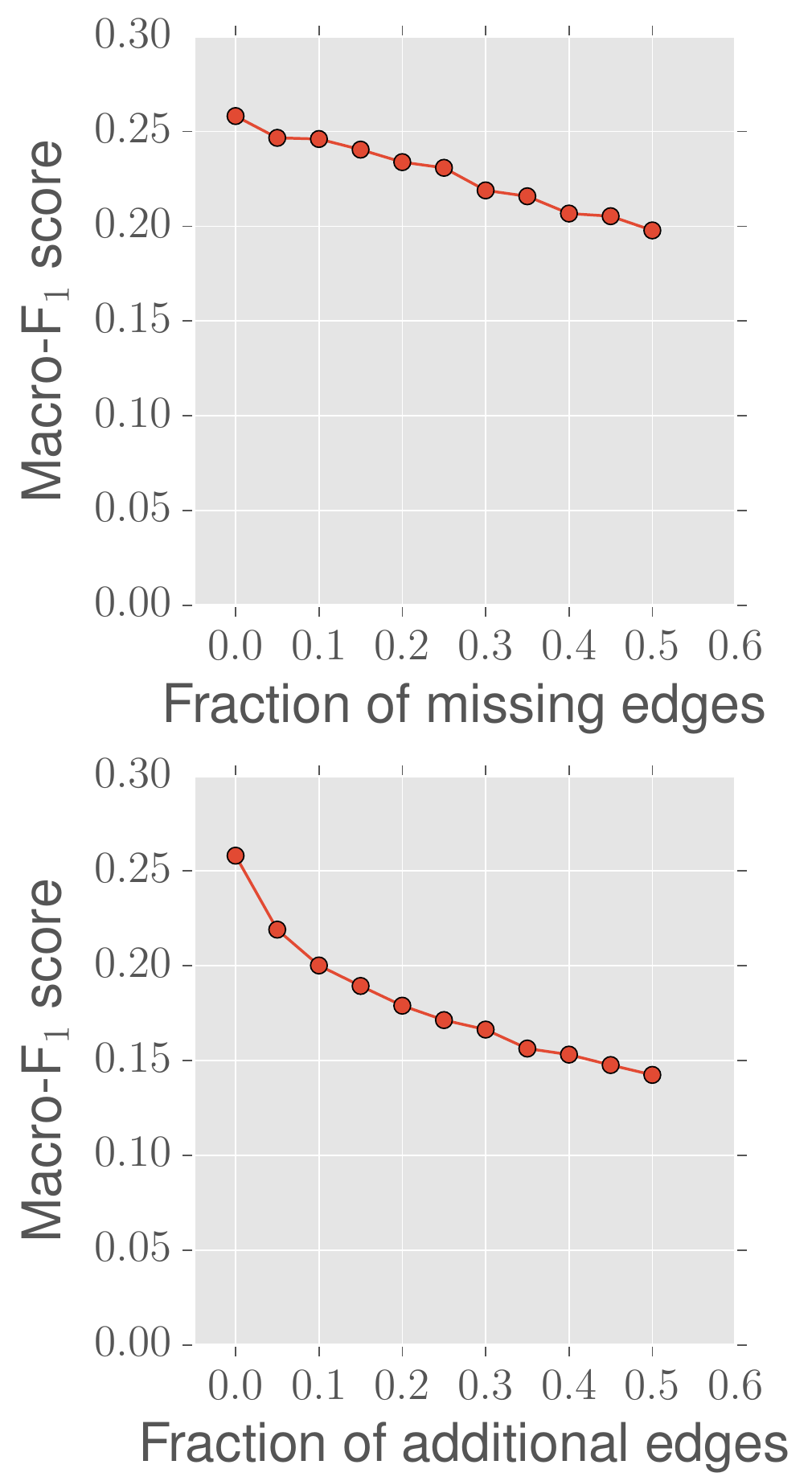}
\caption{}\label{fig:mlc_perturb}
\end{subfigure}
\vspace{-0.2cm}
\caption{(a). Parameter sensitivity (b). Perturbation analysis for multilabel classification on the BlogCatalog network.}
\vspace{-0.4cm}
\end{figure*}

\subsection{Parameter sensitivity} 

The \nodevec algorithm involves a number of parameters and in Figure~\ref{fig:mlc_params}, we examine how the different choices of parameters affect the performance of \nodevec on the BlogCatalog dataset using a 50-50 split between labeled and unlabeled data. Except for the parameter being tested, all other parameters assume default values. The default values for $p$ and $q$ are set to unity.

We measure the Macro-F$_1$ score as a function of parameters $p$ and $q$. The performance of \nodevec improves as the in-out parameter $p$ and the return parameter $q$ decrease. This increase in performance can be based on the homophilic and structural equivalences we expect to see in BlogCatalog. While a low $q$ encourages outward exploration, it is balanced by a low $p$ which ensures that the walk does not go too far from the start node.

We also examine how the number of features $d$ and the node's neighborhood parameters (number of walks $r$, walk length $l$, and neighborhood size $k$) affect the performance. We observe that performance tends to saturate once the dimensions of the representations reaches around 100.
Similarly, we observe that increasing the number and length of walks per source improves performance, which is not surprising since we have a greater overall sampling budget $\mathcal{K}$ to learn representations. Both these parameters have a relatively high impact on the performance of the method. Interestingly, the context size, $k$ also improves performance at the cost of increased optimization time. However, the performance differences are not that large in this case.

\subsection{Perturbation Analysis}
For many real-world networks, we do not have access to accurate information about the network structure. We performed a perturbation study where we analyzed the performance of \nodevec for two imperfect information scenarios related to the edge structure in the BlogCatalog network. In the first scenario, we measure performace as a function of the fraction of missing edges (relative to the full network). The missing edges are chosen randomly, subject to the constraint that the number of connected components in the network remains fixed. As we can see in Figure~\ref{fig:mlc_perturb}(top), the decrease in Macro-F$_1$ score as the fraction of missing edges increases is roughly linear with a small slope. Robustness to missing edges in the network is especially important in cases where the graphs are evolving over time (\eg, citation networks), or where network construction is expensive (\eg, biological networks). 

In the second perturbation setting, we have noisy edges between randomly selected pairs of nodes in the network. As shown in Figure~\ref{fig:mlc_perturb}(bottom), the performance of \nodevec declines slightly faster initially when compared with the setting of missing edges, however, the rate of decrease in Macro-F$_1$ score gradually slows down over time. Again, the robustness of \nodevec to false edges is useful in several situations such as sensor networks where the measurements used for constructing the network are noisy.

\subsection{Scalability}
\begin{figure}[t]
\centering
\includegraphics[width=\columnwidth]{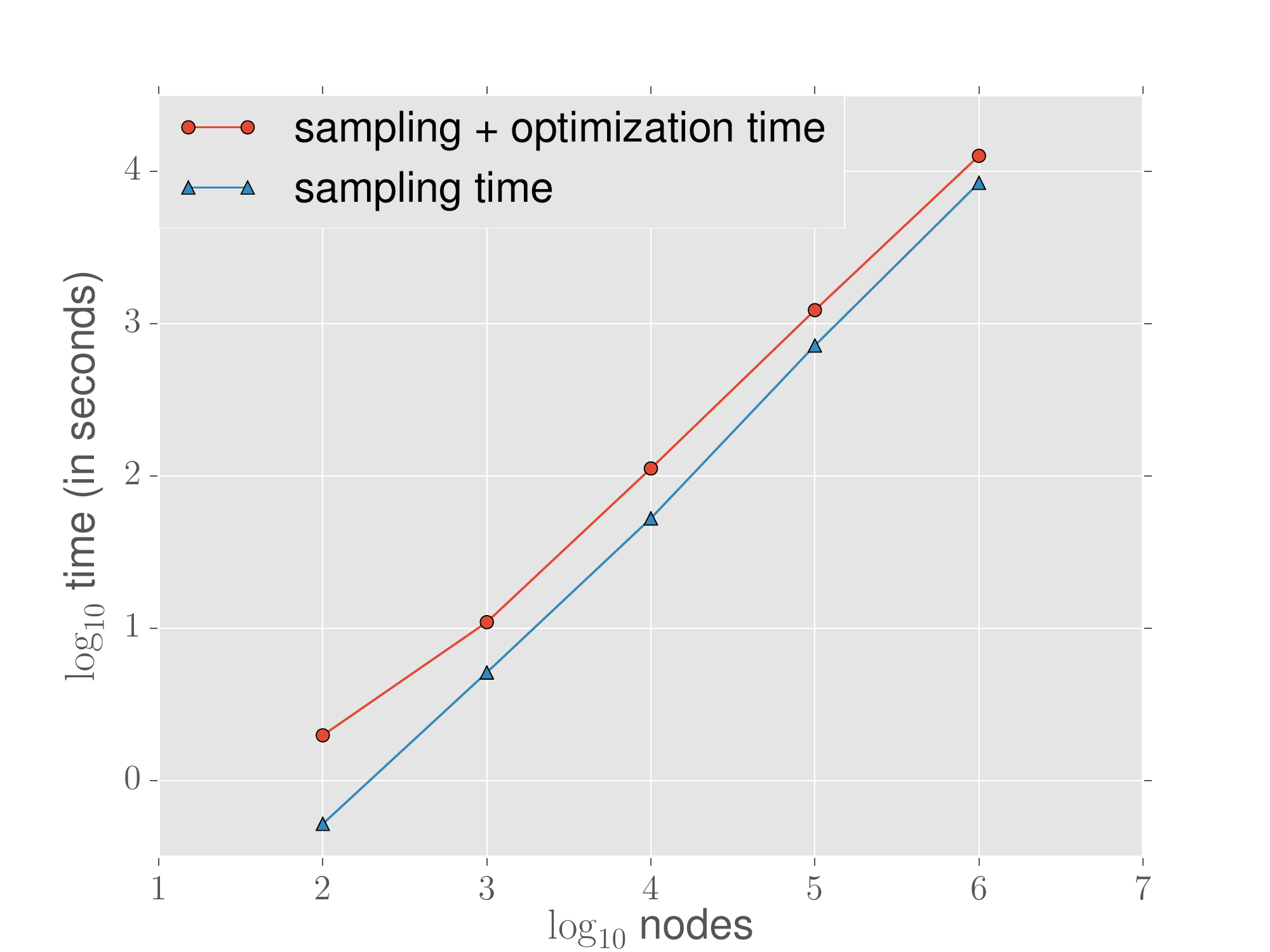}
\vspace{-0.5cm}
\caption{Scalability of \nodevec on Erdos-Renyi graphs with an average degree of 10.}\label{fig:scale}
\vspace{-0.6cm}
\end{figure}

To test for scalability, we learn node representations using \nodevec with default parameter values for Erdos-Renyi graphs with increasing sizes from 100 to 1,000,000 nodes and constant average degree of 10. In Figure~\ref{fig:scale}, we empirically observe that \nodevec scales linearly with increase in number of nodes generating representations for one million nodes in less than four hours. The sampling procedure comprises of preprocessing for computing transition probabilities for our walk (negligibly small) and simulation of random walks. The optimization phase is made efficient using negative sampling~\cite{word2vec2} and asynchronous SGD~\cite{recht-nips2011}. 

Many ideas from prior work serve as useful pointers in making the sampling procedure computationally efficient. We showed how random walks, also used in DeepWalk~\cite{deepwalk}, allow the sampled nodes to be reused as neighborhoods for different source nodes appearing in the walk. Alias sampling allows our walks to generalize to weighted networks, with little preprocessing~\cite{line}. Though we are free to set the search parameters based on the underlying task and domain at no additional cost, learning the best settings of our search parameters adds an overhead. However, as our experiments confirm, this overhead is minimal since \nodevec is semi-supervised and hence, can learn these parameters efficiently with very little labeled data. 


\subsection{Link prediction}
In link prediction, we are given a network with a certain fraction of edges removed, and we would like to predict these missing edges. We generate the labeled dataset of edges as follows: To obtain positive examples, we remove 50\% of edges chosen randomly from the network while ensuring that the residual network obtained after the edge removals is connected, and to generate negative examples, we randomly sample an equal number of node pairs from the network which have no edge connecting them. 

Since none of feature learning algorithms have been previously used for link prediction, we additionally evaluate \nodevec against some popular heuristic scores that achieve good performance in link prediction. The scores we consider are defined in terms of the neighborhood sets of the nodes constituting the pair (see Table~\ref{tab:lp_scores}). We test our benchmarks on the following datasets:

\begin{itemize}[noitemsep,nolistsep]
	\item Facebook~\cite{snapnets}: In the Facebook network, nodes represent users, and edges represent a friendship relation between any two users. The network has 4,039 nodes and 88,234 edges.
	\item Protein-Protein Interactions (PPI)~\cite{biogrid}: In the PPI network for Homo Sapiens, nodes represent proteins, and an edge indicates a biological interaction between a pair of proteins. The network has 19,706 nodes and 390,633 edges.
	\item arXiv ASTRO-PH~\cite{snapnets}: This is a collaboration network generated from papers submitted to the e-print arXiv where nodes represent scientists, and an edge is present between two scientists if they have collaborated in a paper. The network has 18,722 nodes and 198,110 edges.
\end{itemize}

\xhdr{Experimental results} We summarize our results for link prediction in Table~\ref{tab:lp_random}. The best $p$ and $q$ parameter settings for each \nodevec entry are omitted for ease of presentation. A general observation we can draw from the results is that the learned feature representations for node pairs significantly outperform the heuristic benchmark scores with \nodevec achieving the best AUC improvement on 12.6\% on the arXiv dataset over the best performing baseline (Adamic-Adar~\cite{adamicadar}). 

\begin{table}
\centering
\begin{tabular}{l|c}
\textbf{Score} & \textbf{Definition} \\ \hline
Common Neighbors & $\mid \mathcal{N}(u) \cap \mathcal{N}(v) \mid$ \\
Jaccard's Coefficient & $\frac{\mid \mathcal{N}(u) \cap \mathcal{N}(v) \mid}{\mid \mathcal{N}(u) \cup \mathcal{N}(v) \mid}$ \\
Adamic-Adar Score & $\sum_{t \in \mathcal{N}(u) \cap \mathcal{N}(v)} \frac{1}{\log \mid \mathcal{N}(t) \mid}$ \\
Preferential Attachment & $\mid \mathcal{N}(u) \mid \cdot \mid \mathcal{N}(v) \mid$
\end{tabular}
\vspace{-0.2cm}
\caption{Link prediction heuristic scores for node pair $(u, v)$ with immediate neighbor sets $\mathcal{N}(u)$ and $\mathcal{N}(v)$ respectively.}
\label{tab:lp_scores}
\vspace{-0.4cm}
\end{table}

Amongst the feature learning algorithms, \nodevec outperforms both DeepWalk and LINE in all networks with gain up to 3.8\% and 6.5\% respectively in the AUC scores for the best possible choices of the binary operator for each algorithm. When we look at operators individually (Table~\ref{tab:edgeop}), \nodevec outperforms DeepWalk and LINE barring a couple of cases involving the Weighted-L1 and Weighted-L2 operators in which LINE performs better. Overall, the Hadamard operator when used with \nodevec is highly stable and gives the best performance on average across all networks. 

\begin{table}[h]
\centering
	\begin{tabular}{c|l|l|l|l}
	\multicolumn{1}{c}{\textbf{Op}} & \multicolumn{1}{c}{\textbf{Algorithm}} & \multicolumn{3}{c}{\textbf{Dataset}}      \\ 
	         			    & & Facebook     & PPI    & arXiv      \\ \hline
	& Common Neighbors        & 0.8100 & 0.7142 & 0.8153 \\
	& Jaccard's Coefficient   & 0.8880 & 0.7018 & 0.8067 \\
	& Adamic-Adar             & 0.8289 & 0.7126 & 0.8315 \\
	& Pref. Attachment 		  & 0.7137 & 0.6670 & 0.6996 \\ \hline
	& Spectral Clustering     & 0.5960 & 0.6588 & 0.5812 \\
	(a) & DeepWalk            & 0.7238 & 0.6923 & 0.7066 \\ 
	& LINE 			  		  & 0.7029 & 0.6330 & 0.6516 \\
	& \nodevec 		  		  & 0.7266 & 0.7543 & 0.7221 \\ \hline 
	& Spectral Clustering     & 0.6192 & 0.4920 & 0.5740 \\
	(b) & DeepWalk            & \textbf{0.9680} & 0.7441 & 0.9340 \\ 
	& LINE 			  		  & 0.9490 & 0.7249 & 0.8902  \\
	& \nodevec 		  		  & \textbf{0.9680} & \textbf{0.7719} & \textbf{0.9366}\\ \hline 
	& Spectral Clustering     & 0.7200 & 0.6356 & 0.7099\\
	(c) & DeepWalk            & 0.9574 & 0.6026 & 0.8282 \\ 
	& LINE 			  		  & 0.9483 & 0.7024 & 0.8809 \\
	& \nodevec 	      		  & 0.9602 & 0.6292 & 0.8468 \\ \hline 
	& Spectral Clustering     & 0.7107 & 0.6026 & 0.6765 \\
	(d) & DeepWalk            & 0.9584 & 0.6118 & 0.8305 \\ 
	& LINE 			  		  & 0.9460 & 0.7106 & 0.8862 \\
	& \nodevec          	  & 0.9606 & 0.6236 & 0.8477  \\ 
	\end{tabular}
	\vspace{-0.2cm}
	\caption{Area Under Curve (AUC) scores for link prediction. Comparison with popular baselines and embedding based methods bootstapped using binary operators: (a) Average, (b) Hadamard, (c) Weighted-L1, and (d) Weighted-L2
	(See Table~\ref{tab:edgeop} for definitions).
	}\label{tab:lp_random}
	\vspace{-0.4cm}
\end{table}

\section{Discussion and Conclusion}
\label{sec:discussion}

In this paper, we studied feature learning in networks as a search-based optimization problem. This perspective gives us multiple advantages. It can explain classic search strategies on the basis of the exploration-exploitation trade-off. Additionally, it provides a degree of interpretability to the learned representations when applied for a prediction task. For instance, we observed that BFS can explore only limited neighborhoods. This makes BFS suitable for characterizing structural equivalences in network that rely on the immediate local structure of nodes. On the other hand, DFS can freely explore network neighborhoods which is important in discovering homophilous communities at the cost of high variance. 

Both DeepWalk and LINE can be seen as rigid search strategies over networks. DeepWalk~\cite{deepwalk} proposes search using uniform random walks. The obvious limitation with such a strategy is that it gives us no control over the explored neighborhoods. LINE~\cite{line} proposes primarily a breadth-first strategy, sampling nodes and optimizing the likelihood independently over only 1-hop and 2-hop neighbors. The effect of such an exploration is easier to characterize, but it is restrictive and provides no flexibility in exploring nodes at further depths. In contrast, the search strategy in \nodevec is both flexible and controllable exploring network neighborhoods through parameters $p$ and $q$. While these search parameters have intuitive interpretations, we obtain best results on complex networks when we can learn them directly from data. From a practical standpoint, \nodevec is scalable and robust to perturbations.

We showed how extensions of node embeddings to link prediction outperform popular heuristic scores designed specifically for this task. Our method permits additional binary operators beyond those listed in Table~\ref{tab:edgeop}. As a future work, we would like to explore the reasons behind the success of Hadamard operator over others, as well as establish interpretable equivalence notions for edges based on the search parameters. Future extensions of \nodevec could involve networks with special structure such as heterogeneous information networks, networks with explicit domain features for nodes and edges and signed-edge networks. Continuous feature representations are the backbone of many deep learning algorithms, and it would be interesting to use \nodevec representations as building blocks for end-to-end deep learning on graphs.


\xhdr{Acknowledgements}
We are thankful to Austin Benson, Will Hamilton, Rok Sosi\v{c}, Marinka \v{Z}itnik as well as the anonymous reviewers for their helpful comments. This research has been supported in part by NSF
CNS-1010921,     
IIS-1149837,     
NIH BD2K,
ARO MURI, DARPA XDATA,
DARPA SIMPLEX,
Stanford Data Science Initiative,
Boeing,          
Lightspeed,			
SAP,            
and Volkswagen.

\bibliography{refs}
\bibliographystyle{abbrv}


\end{document}